# Satisfiability Modulo Theory based Methodology for Floorplanning in VLSI Circuits

Suchandra Banerjee
mailnmeetsuchandra@gmail.com

Anand Ratna
pacific.anand17@hotmail.com

Suchismita Roy
suchismita27@yahoo.com

Department of Computer Science and Engineering, National Institute of Technology, Durgapur-713209, India

This paper proposes a Satisfiability Modulo Theory based formulation for floorplanning in VLSI circuits. The proposed approach allows a number of fixed blocks to be placed within a layout region without overlapping and at the same time minimizing the area of the layout region. The proposed approach is extended to allow a number of fixed blocks with ability to rotate and flexible blocks (with variable width and height) to be placed within a layout without overlap. Our target in all cases is reduction in area occupied on a chip which is of vital importance in obtaining a good circuit design. Satisfiability Modulo Theory combines the problem of Boolean satisfiability with domains such as convex optimization. Satisfiability Modulo Theory provides a richer modeling language than is possible with pure Boolean SAT formulas. We have conducted our experiments on *MCNC* and *GSRC* benchmark circuits to calculate the total area occupied, amount of deadspace and the total CPU time consumed while placing the blocks without overlapping. The results obtained shows clearly that the amount of dead space or wasted space is reduced if rotation is applied to the blocks.

## 1. INTRODUCTION

Floorplanning is an important step in physical design. Floorplanning helps to provide tentative location of IC building blocks. It is vitally important because it helps to determine size and yield of VLSI chips.

The goal of floorplanning is to optimize a predefined cost function, such as the area of a resulting floorplan given by the minimum bounding rectangle of the layout region. The floorplan area is directly related to the chip silicon cost. The larger the area, the higher the silicon cost. The free space in the floorplan bounding rectangle which is not covered by any module is called white space or dead space. In floorplanning step a design can contain two types of blocks: hard and soft blocks. A hard block is a circuit module with area and aspect ratio fixed. While a soft block has fixed area but variable aspect ratio.

This paper presents a Satisfiability Modulo Theory based approach for floorplanning with hard, soft and rotating basic blocks.

### 1.1 Previous work

Sutanthavibul [5] formulated the floorplanning problem as a mixed integer linear program. The method proposed divides the given problem into various sub problems. Each sub problem is solved using 0-1 mixed integer programming method. The optimal solution for each sub problem is found and the final solution is achieved using successive addition of new elements to an already existing partial solution. Chen [2] have solved the floorplanning problem using particle swarm optimization which is a population based evolutionary algorithm. Their objective is also to minimize the area of hard blocks.

Young [3] in this paper derived a method to handle all kinds of placement constraints at the same time including preplace constraints, range constraints, boundary constraints, alignment constraint, abutement, clustering etc. In this paper, simulated annealing with sequence pair representation is used. Vertical and horizontal constraint graph is used to compute the position of each module. Evangeline and Young [1] used twin binary trees to represent a mosaic floorplan. A mosaic floorplan is a floorplan which has zero spaces. Then a twin binary sequence is constructed. To construct twin binary trees, they start from the module at the lower left corner and travel upward in case of left subtree and to the right in case of right subtree. On reaching the lower left corner of another module a node is inserted in the tree. This process is repeated until all modules are visited. A similar process is done starting from the upper right corner and traveling downward. Then a binary sequence corresponding to the trees is determined.

Tang and Wong [7] discussed fixed frame floorplanning where the problem is addressed by handling alignment constraint which arise in a bus structure. Here, several blocks are aligned in a range (for e.g. bus width) abutting one by one horizontally or vertically. Here, the concept of Sequence Pair Representation was used.

Recent work on floorplanning by Jackey Z Yan and Chris Chu [6] designed an algorithm for soft blocks. Here, a parameter called slack is used which denote the difference in x-coordinate, y-coordinate position between two layouts generated by packing blocks to right, left and top, bottom respectively. The width of the soft blocks on vertical control path are increased and height is decreased correspondingly. Similarly in horizontal control path height is decreased and width is decreased. In [4] it is observed that when two modules are put together, at least one of which has range constraint, the combined super module will also have range constraint. Here, mainly four variables are used to represent the constraint at the right, left, top and bottom. For a particular module with fixed width and height it is constrained to be placed within a bounding rectangle. Then a shape curve is calculated at each internal node till the root which gives us the representation of the possible shapes of the final floorplan. A shape curve is piecewise linear decreasing curve which represents the tradeoff between the height and width of a module as mentioned by Wong [4]. Yan [8] in this paper have induced a nonslicing floorplan by using a compacting technique for slicing floorplan. In comparison to traditional annealing based approaches, one single slicing tree is being considered which is generated by recursive partitioning. The technique here is used to defer the decision on four factors: subfloorplan orientation, subfloorplan order, slice line direction and slicing tree structure. This technique achieves a generalized slicing tree.

### 1.2 Our Contribution

This paper presents an efficient way to handle the floorplanning problem of minimizing the floorplanning region area by using Satisfiability Modulo Theory (SMT) where, the task is to find a satisfying assignment to a set of constraints that minimizes a given objective function. We have successfully placed blocks for the three scenarios as follows:

Case-1: Floorplanning with blocks which have fixed width and height in a layout region.
Case-2: Floorplanning with blocks with fixed width and height but with possibility of change in orientation (blocks may rotate).
Case-3: Floorplanning with blocks with fixed area but flexible width and height.

The constraints which we are using are boundary constraints and non-overlap constraints. The boundary constraints are responsible for maintaining the boundary of the modules within the floorplanning region.

The overlap constraints are responsible for maintaining non-overlapping position of the modules within the region. Including these constraints the objective of area minimization is achieved using SMT.

The paper is organized as follows: Section 2 describes the problem formulation. Section 3 deals with experimental results. Section 4 concludes the paper.

## 2. PROBLEM FORMULATION

The Satisfiability Modulo Theory is the task of finding a satisfying interpretation to a set of variables which satisfies all the constraints. A SMT constraint is expressed as *(assert (<= x 0))*. The *command assert* adds a formula into the solver's internal stack. To use a constant we need to declare it as *(declare-const x Int)*. The formula is satisfiable if there is an interpretation that makes all the asserted formulae true.

### 2.1 Case-1: Floorplanning with blocks which have fixed width and height

Given two rectangular hard blocks $b_i$ and $b_j$ where block $b_i$ is of height $h_i$ and width $w_i$ and block $b_j$ is of height $h_j$ and width $w_j$, our target is to place the modules within a layout region whose bottom left hand corner position is assumed to be (0,0) and top right hand corner position is assumed to be (c,d), such that they do not overlap and at the same time area of the region is minimized. Let $(x_i, y_i)$ and $(x_j, y_j)$ be the left hand corner positions of blocks $b_i$ and $b_j$ respectively. Thus, the constraints for placing the blocks within the floorplanning region are formulated as follows:

*1) Non-Overlap Constraints:* Two blocks i, j are said to be non overlapping if at least one of the following cases is satisfied [5]. To obtain a valid placement of the rectangular modules in the floorplanning region, one of the following constraints must be satisfied:

$$i \text{ to the left of } j \quad x_i + w_i \leq x_j \quad (1)$$
$$i \text{ below } j \quad y_i + h_i \leq y_j \quad (2)$$
$$i \text{ to the right of } j \quad x_i - w_j \geq x_j \quad (3)$$
$$i \text{ above } j \quad y_i - h_j \geq y_j \quad (4)$$

In order to ensure that of the four inequalities, at least one holds true, two additional Boolean variables $x_{ij}$ and $y_{ij}$ are introduced. These variables can take only values 0 or 1.

$$x_i + w_i \leq x_j + W(x_{ij} + y_{ij}) \quad (5)$$
$$x_i - w_j \geq x_j - W(1 - x_{ij} + y_{ij}) \quad (6)$$
$$y_i + h_i \leq y_j + H(1 + x_{ij} - y_{ij}) \quad (7)$$
$$y_i - h_j \geq y_j - H(2 - x_{ij} - y_{ij}) \quad (8)$$

Here W, H are maximum allowable width and height of the layout region and $(x_i, y_i)$ and $(x_j, y_j)$ are the bottom left hand corner positions of blocks $b_i$ and $b_j$ respectively. The width and height of the blocks $b_i$ and $b_j$ are $w_i$, $h_i$ and $w_j$, $h_j$ as mentioned earlier.

*2.) Boundary Constraints:* To ensure that the blocks lie within boundary of the floorplanning region the following constraints are used:

$$x_i + w_i \leq c \quad (9)$$
$$y_i + h_i \leq d \quad (10)$$
$$x_j + w_j \leq c \quad (11)$$
$$x_i + w_i \leq d \quad (12)$$

*3) The objective function:* Since the target of the problem is to minimize the area of the floorplanning region, the objective function is formulated as:

minimize:   c * d

where (c,d) is the top right hand corner position of the floorplanning region and the lower left hand corner position is considered to be (0,0).

Example: As an example let us consider 5 rigid or hard blocks with fixed width and height (10,15), (8,6), (9,5), (9,7), (7,8) respectively. To place these blocks we have used the non-overlapping and boundary constraints as mentioned above. Our objective is to place the blocks within the layout region at the same time minimizing the total area. The result after placement of the blocks after applying our technique is represented in the following Fig. 1(a).

*Algorithm:*

Input: Set of n blocks with fixed width and height.
Output: Floorplanning region with minimized area.

1: Calculate W (sum of width of the blocks).
2: Calculate H (sum of height of the blocks).
3: for i = 1 to n do
4:     Add boundary constraints
5: end for
6: for i =1 to n do
7:    for j = 1 to n do
8:     if i != j
9:      Add non-overlapping constraints for
         block i and j
10:    end if
11:   end for
12: end for
13: Minimize area of floorplanning region.
14: End

## 2.2 Case 2: Floorplanning with blocks which can rotate.

Given two rectangular hard blocks $b_i$ and $b_j$ where $b_i$ is of height $h_i$ and width $w_i$ and $b_j$ is of height $h_j$ and width $w_j$, our target is to place the blocks within a floorplanning region whose bottom left hand corner position is assumed to be (0,0) and top right hand corner position is assumed to be (c,d), such that they do not overlap and at the same time area of the layout is minimized. For the second category blocks have additional condition of rotation. The blocks can rotate by $90^0$.

*1) Non-Overlap Constraints*
Two blocks i, j are said to be non overlapping if at least one of the following cases is satisfied [5]. Here, one additional Boolean variable $z_i$ is used to enable change in orientation of the block where $z_i = 1$ when the module is rotated by $90^0$ and $z_i = 0$ when placed in its initial orientation.

$$x_i + z_i h_i + (1- z_i) w_i \leq x_j + M(x_{ij} + y_{ij}) \quad (13)$$
$$x_i - z_j h_j - (1- z_j) w_j \geq x_j - M(1- x_{ij} + y_{ij}) \quad (14)$$
$$y_i + z_i w_i + (1- z_i) h_i \leq y_j + M(1+ x_{ij} - y_{ij}) \quad (15)$$
$$y_i - z_j w_j - (1- z_j) h_j \geq y_j - M(2- x_{ij} - y_{ij}) \quad (16)$$

Here, M = maximum of W or H mentioned earlier.

*2) Boundary Constraints*
To ensure that the block lies within boundary of the floorplanning region the following constraints are used:

$$x_i + (1- z_i) w_i + z_i h_i \leq c \quad (17)$$
$$y_i + (1- z_i) w_i + z_i h_i \leq d \quad (18)$$

*3) The objective function*
In this case also the target of the problem is to minimize the area of the floorplanning region, the objective function is formulated as:

minimize: c * d

As an example we have considered the same 5 rigid or hard blocks with fixed width and height taken earlier. After applying our method the result improved as depicted in Fig. 1(b).

## 2.3 Case-3: Floorplanning with soft blocks

Given two blocks of fixed area $A_i$ and $A_j$ respectively. The width and height of the blocks ($w_i$, $h_i$) and ($w_j$, $h_j$) may vary. Our target is to place the blocks within a floorplanning region whose bottom left hand corner position is assumed to be (0,0) and top right hand corner position is assumed to be (c,d), such that they do not overlap and at the same time area of the layout is minimized.

## 1) Constraints

The non-overlap constraints and boundary constraints are used which are same as in Case-2. Apart from these constraints additional constraint are used is as follows:

$$w_i * h_i = A_i \quad (19)$$

To maintain the shape of the blocks we have used an aspect ratio range of [0.1, 10].

As an example we have considered the same 5 flexible blocks with fixed area of 150, 48, 45, 63 and 56 units respectively. After applying our method the resulting placement of the blocks are depicted in Fig. 1(c).

## 3. EXPERIMENTAL RESULTS

We have formulated the proposed floorplanning technique as a SMT problem and implemented it using C and run it on the standard GSRC benchmark circuits and the MCNC benchmark circuits. The formulated problem has been solved with z3 as the underlying SMT solver.

All experiments have been performed on a machine with 128 GB memory and operating system Cent-OS. Satisfiability Modulo Theory technique is used to find a floorplan with no overlapping and occupying minimal area.

In Table I, we have given 5 MCNC benchmark circuits whose total area is calculated in $mm^2$. In Table II, we have given the resultant area of the layout region for 3 GSRC benchmark circuits.

Table I
AREA OF MCNC BENCHMARK CIRCUITS

| Circuit | No. of blocks | Case1($mm^2$) | Case2($mm^2$) | Case3($mm^2$) |
|---|---|---|---|---|
| *apte* | 9 | 48.05 | 48.12 | 50.24 |
| *xerox* | 10 | 20.44 | 20.01 | 21.04 |
| *hp* | 11 | 9.36 | 9.26 | 10.44 |
| *ami33* | 33 | 1.37 | 1.28 | 1.30 |
| *ami49* | 49 | 42.37 | 40.04 | 51.27 |

After experimentation we have found that the best results are obtained by rotating the blocks so we extended our work on larger GSRC benchmark circuits consisting of 100, 200 and 300 blocks. To handle the larger circuits, it was necessary to break up the circuits into spatial clusters. We have also found out the percentage of deadspace and CPU time for all of the following cases depicted in the Table III given below.

We have also observed from the Table III and Table IV that the amount of deadspace obtained after applying rotation (Case 2) of blocks is reduced to significantly.

Table II
AREA OF GSRC BENCHMARK CIRCUITS

| Circuit | No. of blocks | Case1($mm^2$) | Case2($mm^2$) | Case3($mm^2$) |
|---|---|---|---|---|
| *n10* | 10 | 0.231 | 0.226 | 0.240 |
| *n30* | 30 | 0.235 | 0.218 | 0.257 |
| *n50* | 50 | 0.218 | 0.211 | 0.220 |

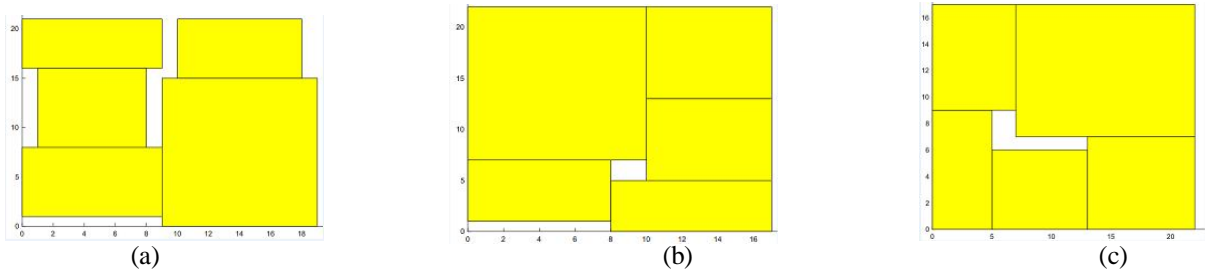

Fig. 1.Floorplanning with a)fixed blocks b) fixed blocks with rotation c) flexible blocks for example given in Section 2.

In Table V, we have given a comparison of our result with the results given in [2] for benchmark circuits, expressed as area occupied by the blocks. It can be seen that our technique gives better result in MCNC circuits *ami33*, *xerox* and GSRC benchmark circuits *n30* and *n50*.

Fig. 2, 3 and 4 depict the resultant placement of the blocks for *GSRC* and MCNC benchmark circuits using rotation. Fig. 2 depicts the resultant placement of MCNC benchmark *ami33* with and without using rotation. Fig. 3 depicts the resultant placement of GSRC benchmark of 300 blocks without using rotation while Fig. 4 depicts the resultant placement of 300 blocks using rotation.

It has been observed after performing the experiments that the results improved after applying rotation to the fixed blocks while no great improvement in result was being observed for flexible blocks. It can be seen that the dead space in the layouts is primarily owing to the straight forward spatial clustering process used to handle larger circuits. Experimenting with more sophisticated clustering techniques with the target of reducing dead space in the layout is currently work in progress.

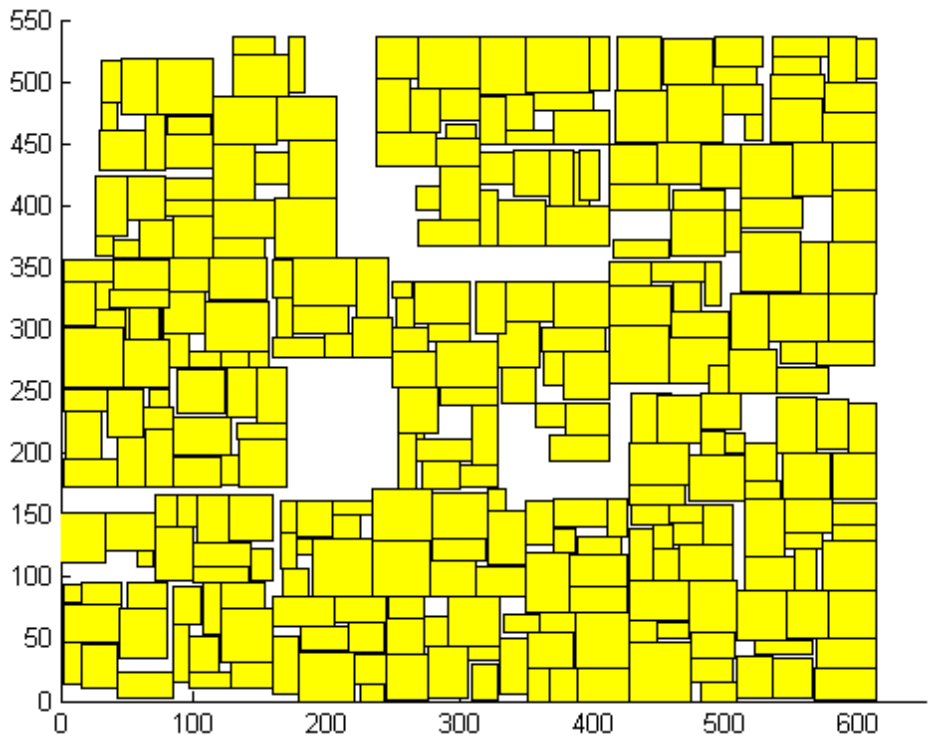

Fig. 3 Floorplanning of GSRC benchmark circuit n300 without using rotation

Table III
COMPARISON OF CASE 1 AND CASE 2 FOR GSRC BENCHMARK CIRCUITS

| Circuit | Case 1 (without rotation) | | | Case 2 (with rotation) | | |
|---|---|---|---|---|---|---|
| | Area(mm$^2$) | % dead space | CPU time (sec) | Area(mm$^2$) | %dead space | CPU time (sec) |
| n30 | 0.235 | 11.86 | 64.19 | 0.218 | 4.7 | 1220.11 |
| n 50 | 0.218 | 8.79 | 219.38 | 0.211 | 5.9 | 4386.68 |
| n 50b | 0.224 | 9.46 | 569.17 | 0.210 | 3.35 | 45474.2 |
| n 50c | 0.242 | 16.8 | 362.35 | 0.242 | 16.8 | 4066.42 |
| n 100 | 0.226 | 20.68 | 314.67 | 0.192 | 6.6 | 8745.29 |
| n 100b | 0.204 | 21.5 | 269.89 | 0.182 | 12.2 | 10177.33 |
| n 100c | 0.221 | 22.35 | 221.14 | 0.197 | 12.7 | 6236 |
| n 200 | 0.229 | 23.4 | 483.72 | 0.205 | 14.4 | 10096 |
| n 200b | 0.235 | 25.8 | 934 | 0.196 | 11.3 | 11549 |
| n 200c | 0.215 | 20.8 | 482 | 0.194 | 12.6 | 9255 |
| n 300 | 0.329 | 20.70 | 931.72 | 0.316 | 13.5 | 26777 |

Table IV
COMPARISON OF CASE 1 AND CASE 2 FOR MCNC BENCHMARK CIRCUITS

| Circuit | Case 1 (without rotation) | | | Case 2 (with rotation) | | |
|---|---|---|---|---|---|---|
| | Area(mm$^2$) | % dead space | CPU time (sec) | Area(mm$^2$) | % dead space | CPU time (sec) |
| apte | 48.05 | 3.2 | 4260 | 48.12 | 3.2 | 24280 |
| xerox | 20.44 | 5.53 | 3019.26 | 20.11 | 3.3 | 29282 |
| hp | 9.36 | 5.6 | 323.6 | 9.26 | 4.6 | 2013.37 |
| ami 33 | 1.37 | 15.6 | 428.13 | 1.28 | 9.9 | 29211 |
| ami 49 | 42.37 | 16.3 | 155.41 | 40.04 | 11.4 | 11310 |

Table V
COMPARISON OF THE LAYOUT AREA OBTAINED IN [2] AND USING OUR PROPOSED TECHNIQUE
(CASE 2: WITH ROTATION)

| Circuit | Area in [2] mm$^2$ | Our result (using rotation) mm$^2$ |
|---|---|---|
| xerox | 20.38 | 20.01 |
| ami33 | 1.29 | 1.28 |
| n30 | 0.234 | 0.218 |
| n50 | 0.222 | 0.211 |

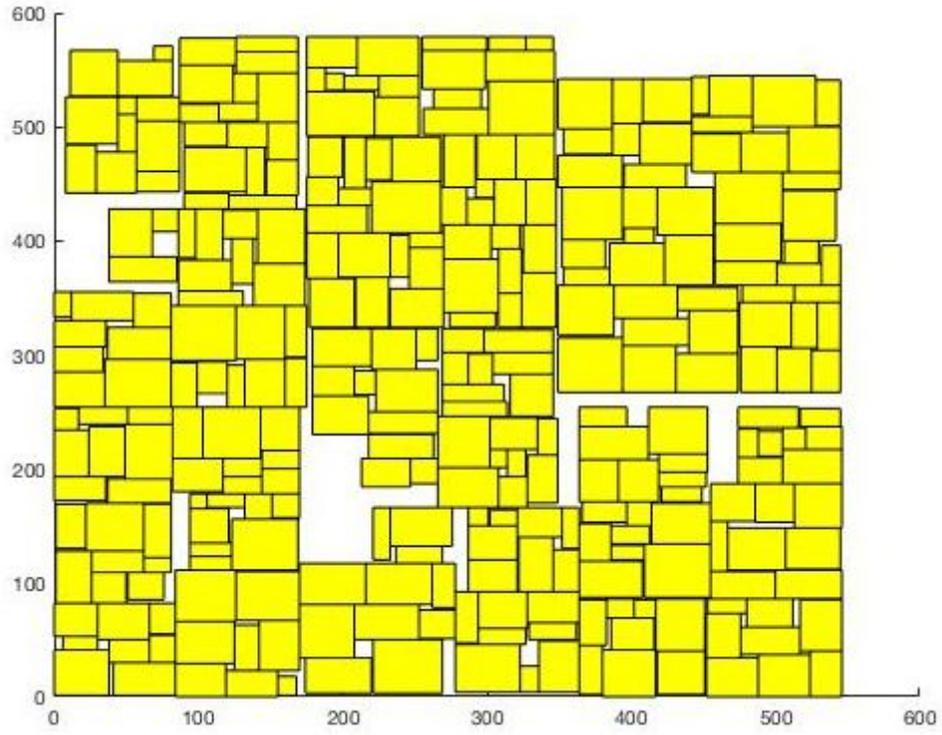

Fig. 4. GSRC benchmark circuit n300 after floorplanning using rotation

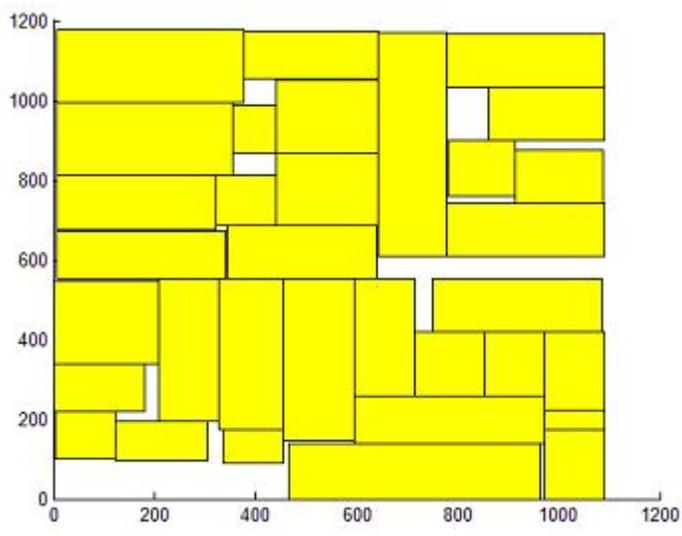
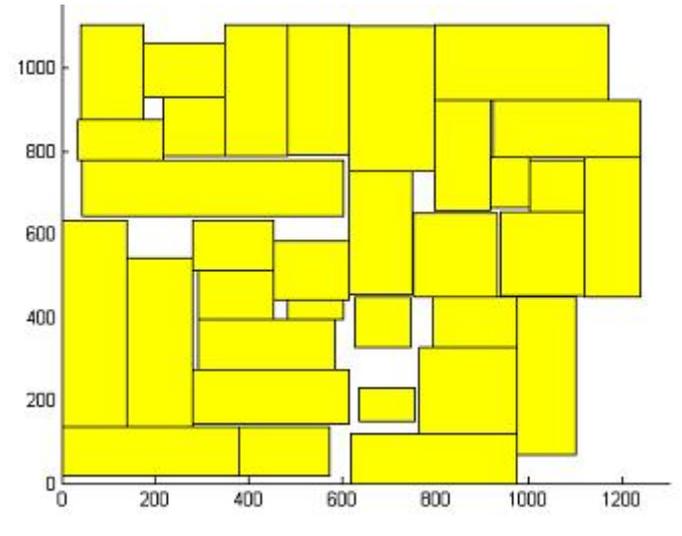

a)                                                b)

Fig. 2.    MCNC benchmark ami33 after floorplanning    a) using rotation     b) without rotation

## 4. CONCLUSION

The proposed technique performs an important step in placing fixed blocks on a floorplanning region without overlapping using the concept of Satisfiability Modulo Theory. Our technique is useful in floorplanning with realistic blocks whose width and height are fixed. The objective is to minimize the area of the floorplanning region so that chip area is best utilized. A SMT based solution provides an optimal solution to the problem after exploring all possibilities.

**Acknowledgments:** This research is supported by the Department of Electronics and Information Technology, Ministry of Communications and IT, Government of India under the Visvesvaraya PhD Scheme administered by Media Lab Asia.

## References

[1] Young, Evangeline F. Y. and Chu, Chris C. N. and Shen, Cien, "Twin Binary Sequences: A Non-redundant Representation for General Non-slicing Floorplan," ISPD '02, pages 196-201.

[2] Chen Guolong, Guo Wenzhong and Chen Yuzhong, "A PSO-based intelligent decision algorithm for VLSI floorplanning," Soft Computing journal, 2010, pp. 1329-1337.

[3] Young, Evangeline F. Y. and Chu, Chris C. N. and Ho, M. L., "Placement Constraints in Floorplan Design," IEEE Trans. Very Large Scale Integr. Syst. , July 2004, vol. 12, pp. 735--745.

[4] Young, F. Y. and Wong, D. F. and Yang, H. H. , "Slicing Floorplans with Range Constraint," Trans. Comp. Aided Des. Integ. Cir. Sys. , November 2006, vol. 19, pp. 272--278.

[5] Sutanthavibul, Suphachai and Shragowitz, Eugene and Rosen, J. Ben, "An Analytical Approach to Floorplan Design and Optimization," DAC '90, pp. 187-192.

[6] Yan, Jackey Z. and Chu Chris, "Optimal Slack-driven Block Shaping Algorithm in Fixed-outline Floorplanning," ISPD '12, pp. 179-186.

[7] Tang, Xiaoping and Wong, D. F. , "Floorplanning with Alignment and Performance Constraints," DAC '02, pp.848-853.

[8] Yan Jackey Z. and Chu Chris, "DeFer: Deferred Decision Making Enabled Fixed-outline Floorplanning Algorithm," Trans. Comp.-Aided Des. Integ. Cir. Sys. , March 2010, vol. 29, pages 367-381.

[9] Sherwani Naveed A. , Algorithms for VLSI Physical Design Automation, Kluwer Academic Publishers, Norwell, MA, USA, 1995.

[10] Moura, Leonardo and BjØrner, Nikolaj, "Satisfiability Modulo Theories: An Appetizer," 2009, Springer-Verlag, pages 23-36.